# Awareness and use of quantitative decision-making methods in pharmaceutical development

Guido Thömmes[1], Martin Oliver Sailer[2], Nicolas Bonnet[3], Alex Carlton[4], Juan J. Abellan[4], and Veronique Robert[5], on behalf of the European Special Interest Group on Quantitative Decision-Making

[1]Grünenthal GmbH, Aachen, Germany. [2]Boehringer Ingelheim Pharma GmbH & Co. KG, Biberach, Germany. [3]Sanofi, Montpellier, France. [4]GlaxoSmithKline, Brentford, Middlesex, UK. [5]Servier, Suresnes, France.

## Abstract

The pharmaceutical industry has experienced increasing costs and sustained high attrition rates in drug development over the last years. One proposal that addresses this challenge from a statistical perspective is the use of quantitative decision-making (QDM) methods to support a data-driven, objective appraisal of the evidence that forms the basis of decisions at different development levels. Growing awareness among statistical leaders in the industry has led to the creation of the European EFSPI/PSI special interest group (ESIG) on quantitative decision making to share experiences, collect best practices, and promote the use of QDM. In this paper, we introduce key components of QDM and present examples of QDM methods on trial, program, and portfolio level. The ESIG created a questionnaire to learn how and to what extent QDM methods are currently used in the different development phases. We present the main questionnaire findings, and we show where QDM is already used today but also where areas for future improvement can be identified. In particular, statisticians should increase their visibility, involvement, and leadership in cross-functional decision-making.

## 1. Introduction

In recent years, the pharmaceutical industry has faced the challenge of increasing costs and constantly high attrition rates which tend to make its business model less sustainable in the future. The increase of capitalized cost invested per approved new drug is nowadays estimated around 2.5 billion US dollar (DiMasi et al. 2016) and has resulted in the pharmaceutical industry changing its drug development model into a "Quick win, fast fail" approach to improve R&D productivity (Paul et al. 2010). In this approach, the proof-of-concept (PoC) trial is a key component where the false-negative decision should be carefully controlled. A PoC trial is the earliest point in the drug development process at which the weight of evidence suggests that it is "reasonably likely" that the key attributes for success are present and the key causes of failure are absent (Cartwright et al. 2010). The use of PoC trials has reduced attrition rates in recent years but they continue to remain high (Wong and Siah 2019).

This suggests the important role appropriate methods for evidence assessment can play to make informed decisions which appropriately take uncertainty into account. To address this challenge and to contribute to possible mitigations, statisticians have proposed the use of quantitative decision-making (QDM) methods supporting a data-driven, objective appraisal of the evidence that forms the basis of decisions at different stages of drug development. Evidence-based quantitative methods in a decision-making framework provide an integrated assessment of the available evidence and its associated uncertainty, thus enabling rational decisions (Ashby and Smith 2000). Statistical thinking helps to make better decisions by increasing awareness of the risks in the presence of uncertainty and by quantifying the risks involved in the decisions. In this sense, statisticians should embrace QDM methods to better support decisions and eventually to increase the chances of successful drug development.

QDM methods can be applied at trial, program, and portfolio level, and various authors have made proposals in the last years (O'Hagan et al. 2005, Lalonde et al. 2007, Frewer et al. 2016, Antonijevic 2015). The growing interest in this topic from the pharmaceutical industry in recent years has also led to the creation of a cross-industry special interest group sponsored by the European Federation of Statisticians in the Pharmaceutic Industry (EFSPI) and Statisticians in the Pharmaceutical Industry (PSI). The European Quantitative Decision-making Special Interest Group (QDM ESIG) was formed in 2017. It is a group of statisticians from industry and academia with experience and interests in statistical methods for quantitative decision-making in drug development. The objectives of the ESIG are to share experiences, collect best practices and promote the use of QDM methods in drug development. Specifically, the group organizes events bringing together statisticians from the pharmaceutical industry and academia to exchange about recent advances in statistical methods for quantitative decision-making in drug development.

Despite the growing general awareness of QDM methods, it has been less clear to what extent and how these methods are currently used in practice in the pharmaceutical industry. Therefore, the ESIG created a questionnaire that was sent to colleagues in the pharmaceutical industry to learn about the current status of the use of QDM methods. In Section 2 we introduce a background and key components of QDM. Motivating examples are provided in Section 3. Section 4 presents the results from the questionnaire on the use of QDM methods conducted by the QDM ESIG followed by a discussion in Section 5 and conclusions in Section 6.

## 2. Background

Decision-making is one of the main tasks of drug development at different levels, from individual trials to programs and to entire portfolios. Statistics can play an important role to support evidence-based decisions by providing principled methods for the appraisal of the available data, e.g., using quantitative decision-making frameworks. The use of statistical methods for decision-making has been promoted for a long time (Lalonde et al. 2007). This section motivates the use of statistical decision-making methods and provides the background for the creation of ESIG on QDM in 2017.

### 2.1 Decision-making at trial, program and portfolio level

In the context of drug development, many decisions have to be made at clinical trial, program and portfolio level. When planning a trial, choosing the appropriate design and decision framework is particularly important. In addition to conventional power calculations, simulation techniques can be employed to evaluate the performance of potential designs in order to understand the operating characteristics, e.g., probability of false-positive, false-negative and of making the right decision, of each design (Lalonde et al. 2007). During the trial, modeling patient recruitment may also be used to support operational decisions (drug supply, opening of centers, etc.), and interim analyses can allow for decision-making whether to stop the trial early or to continue the trial. At the end of the clinical trial, the outcome may lead to a decision to stop developing the drug or to continue to the next stage of development. Until now, in a frequentist framework, traditional tools such as significance tests and estimations were mostly used for end of trial decision-making.

At the program level, there are many decision points along the development path. In this long journey, data needs to be regularly evaluated to decide if the development should be continued. Formal clinical milestones, such as the end of phase I, phase II, or phase III provide useful decision-points to critically evaluate the accumulating data and make important Go/NoGo decisions. A correct decision is important to avoid wasting resources on non-promising compounds and to avoid terminating valuable drugs prematurely (Chuang-Stein et al. 2011).

To allow informed decision-making and to increase the chance of correct decisions at each step of product development is paramount. Ultimately decision makers in pharmaceutical companies have to make decisions considering several products at the portfolio level. The structure of the product pipeline is driven by the cost of development of the drugs, the likelihood of them reaching regulatory approval, and finally the expected profitability. In the past, decisions at portfolio level were commonly made on the basis of intuition, human judgment and prior experience (Skrepnek and Sarnowski 2007). Today there is increasing awareness that more formal methods of analyzing drug pipeline portfolios are essential to make more informed decisions (Jekunen 2014), and these quantitative approaches can support the decision-making process and supplement existing heuristic approaches for portfolio management (Patel et al., 2020).

### 2.2 Evidence-based methods in a decision-making framework

The iterative process of collecting and updating the evidence helps us make decisions using relevant metrics (Chuang-Stein et al. 2011). At the same time, there is a need to take this best available evidence, and associated uncertainty, and combine it with judgments about risks, benefits and costs in order to make decisions (Ashby and Smith 2000). Therefore, we consider that QDM methods aim to support an optimal and informed choice, at a given time point, based on the cumulative available information integrating preferences of the decision-maker(s).

As evidence-based methods, from Ashby's considerations, QDM methods should require an integrated assessment of the available evidence and associated uncertainty which must be taken into account in order to make rational decisions. This decision-making should also be based on pre-specified criteria, in a decision-making framework, as proposed by several authors previously.

Around 15 years ago, Lalonde introduced quantitative decision-criteria as rules which use the estimation of treatment effect and its distribution, i.e., the available evidence and associated uncertainties, to decide the course of action for a compound, especially after completing a trial. The decision-making rules are often based on statistical significance, a pre-specified target effect magnitude (target value, TV) and/or a pre-specified lower reference value (LRV), for example defined as the treatment effect of a competitor known to have mild efficacy (Lalonde et al. 2007, Walley et al. 2015, Frewer et al. 2016).

In addition, when deciding whether to move into a confirmatory stage, a good metric would be the probability that there will be a successful confirmatory trial outcome (Chuang-Stein et al. 2011). In this case, existing information is incorporated into future trial planning, and assurance, i.e., expected success probability averaged over a prior probability distribution for the treatment effect, is used to quantify the probability of success of this future study (O'Hagan et al. 2005). The implementation of predictive probability of success (PPoS) as a routine metric to support decision-making has grown across the industry in the recent years (Crisp et al. 2018).

In the assurance setting, eliciting prior knowledge plays a critical role. In principle, historical data can provide a prior probability distribution, e.g. for a treatment effect. Also, formal expert elicitation methods, i.e. the process of extracting the expert's information (knowledge or opinion) related to quantities or events of interest that are uncertain, can be used to form the prior (Dallow et al. 2018).

To set up such quantitative tools requires interaction between the members of a multi-disciplinary team where the statistician plays a key role. Statisticians, the scientists of uncertainty, are perfectly placed to quantitatively inform the decision-making process, supporting it with tools like decision analysis, optimization, modeling, and simulation (Patel et al. 2020). With their understanding of the strengths and limitations of data, statisticians should proactively initiate and drive the discussions around quantitative decision making.

### 3. Examples of QDM in pharmaceutical development

In this section we present three examples to illustrate how QDM can be applied. First, an oncology trial illustrates how a decision-making framework can facilitate Go/NoGo decisions in phase II. Secondly, the use of assurance calculations to quantify the probability for a successful trial result is shown for a cardiovascular outcome trial in phase IIIb. And thirdly, an example is given showing how a quantitative approach can help maximize the value of a late-stage development portfolio. For a discussion of the statistical methodology of these QDM methods we refer to a companion paper by Abellan et al. (submitted).

#### 3.1. Go/NoGo decision making criteria for a phase II Oncology trial

Frewer et al. (2016) presented a randomized phase II oncology study to investigate whether an experimental drug extends the primary endpoint progression-free survival (PFS) compared to control. Based on available literature, a TPP was set up with a base case increase of the median PFS from 10 to 14.6 months (hazard ratio HR=0.68, target value TV) and a downside case increase of the median PFS from 10 to 12 months (HR=0.83, lower reference value LRV).

For randomized phase II trials, hypothesis testing is typically used to assess the objective, often with higher type I error rates and possibly lower power compared to a pivotal trial in order to allow for reduced sample sizes. Frewer et al. (2016), however, used a three-region decision-making approach:
- Go if observed HR < 0.72
- Consider if observed HR > 0.72 and < 0.86
- Stop if observed HR >= 0.86.

It may be noted that the LRV and TV thresholds differ in general from the limits of the Go and Stop zone. LRV and TV refer to the true value of the parameter PFS in the population, whereas the Go and Stop limits refer to the parameter estimate observed in the trial. The limits for the PFS point estimate are derived from the population thresholds by taking pre-specified False-Go (type I error) and False-Stop (type II error) rates into account. Rather than controlling the type I error rate for the clinically meaningless null hypothesis of no PFS benefit, the error rates to be controlled for were defined relative to the clinically and commercially meaningful TV and LRV. If the true HR equalled the TV, then the probability for a Go or Consider was to equal 90%. However, 90% power to obtain a Go alone was not required. If the true HR equalled the borderline clinically relevant LRV, the probability for a Go should have been be at most 20%.

If the data suggested a moderate benefit, the team was to consider further action based on additional information, e.g. using other endpoints. In the end, all relevant information will have been used to come to a decision whether to continue the development project or not. The three-region approach was a simple way to introduce multiple endpoints into the decision-making algorithm that provided the clinical team with some flexibility. In order to limit arbitrary or inconclusive outcomes the design was calibrated such that the probability to end up in the Consider zone was at most 30%. Owing to the introduction of the Consider zone and the above requirements for the operating characteristics, a considerably smaller sample size was achieved compared to a design with only Go or NoGo decision regions. This came at the price of reduced probabilities to achieve a Go for a base case drug and to achieve a NoGo for an insufficient drug. Even with the review of additional data in the consider zone there was an increased risk to eventually come to the wrong conclusion about further development. In the example, the risks associated with these operating characteristics were considered acceptable.

### 3.2 Leveraging assurance for phase III trial design

Crisp et al. (2018) present an example of a cardiovascular outcomes trial in phase IIIb. The study was initially powered to show non-inferiority. Superiority was included as a secondary objective. In addition to usual conditional operating characteristics like power and type I error, the team also considered the PoS of the trial for the final trial design.

During the trial, promising external data emerged that led to discussions about an increase of the originally planned number of events for the superiority analysis. An amended design was considered such that non-inferiority was tested at an interim analysis, followed by a final analysis for superiority at a later final analysis timepoint. An interim futility analysis was introduced to limit the risk of increasing the sample size only to end up failing to show superiority.

The study team aided the decision-making on the amendment by showing not only the overall PoS for superiority of the trial but also the probability of a Go at interim and the conditional PoS of showing superiority conditional on a Go at interim.

PoS was determined via assurance. The prior distribution for the treatment effect was obtained via a modification of the prior elicitation process described in Dallow et al. (2017). Priors were elicited from individual experts and averaged to arrive at a consensus prior to be used for the PoS calculation.

Eventually, the inclusion of a futility analysis and subsequent increase in sample size increased the PoS from 31% to 40%. The conditional PoS after passing the interim was very high (80%) and thus could demonstrate to decision makers the value of the interim analysis. The PoS determined by assurance was able to guide the final design of the trial.

### 3.3 Portfolio optimization

Several methods have been presented in the literature for the management of a pharmaceutical portfolio in terms of selecting which studies should be conducted. Many methods aim to maximize the expected net present value (ENPV). The ENPV is indeed one way to capture the value of a pharmaceutical product through the difference between the present value of future returns from an investment and the amount of investment itself, weighted by its PoS.

Patel and Ankolekar (2015) and Patel et al. (2020) focused on approaches that explicitly model the relationship between trial designs and performance criteria like PoS and ENPV. They use a model for dynamically optimizing budget allocation for phase III drug development portfolios that incorporates uncertainty in the pipeline. The objective of the model is to maximize ENPV of a portfolio over a planning horizon by determining the optimal design of phase III trials for a given budget. The model provides an optimal policy that specifies the optimal design for each drug for every possible scenario of availability of future drugs for phase III trials. It optimizes the trade-off between committing budget allocations to drugs available for phase III funding at any point in time and preserving budget for drugs in the development pipeline that will need funding in the future.

In an illustrative example, the authors show how the optimization model can be used to decide on the best budget level to meet a target Return on Investment (ROI), defined as ENPV/Budget. In this example, the ROI has almost doubled by using an optimal dynamic strategy to allocate phase III budget instead of a static strategy that maximizes portfolio ENPV by maximizing ENPV for each drug individually.

### 4. Current awareness and the use of QDM methods

As described in Section 2, there are good reasons for using quantitative methods to improve decision-making in pharmaceutical development, and in recent years a number of approaches have been proposed mainly on a trial and program level but also on the portfolio level. Section 3 highlights a few successful published examples. However, it is less clear to what extent and how these methods are currently used in practice in the pharmaceutical industry. Therefore, the EFSPI/PSI special interest group created a questionnaire to fill this gap. The questionnaire was sent to fellow statisticians as well as to non-statisticians, including decision makers, working at different stages of the clinical development in the pharmaceutical industry. The objectives were two-fold. First, to elicit the current status of the use of QDM methods and, secondly, to understand the needs of the customers and consumers of QDM results.

The survey was conducted in Europe jointly by EFPSI and PSI in January and February 2019. An online questionnaire was sent to members of the two statistical organizations and recipients were asked to forward it to interested colleagues, in particular also to non-statisticians and decision makers. The resulting answers received from the participants are an anonymous, non-representative sample but should nevertheless give an impression how QDM methods are currently used in practice by working statisticians and their non-statistical colleagues.

In total 101 responses were received, 54 from statisticians and 47 from non-statisticians, the latter including 15 decision makers, i.e., members of committees where key development decisions are made. Responses were received mainly on the study and the program level, reflecting perhaps the fact that statistical methods for trial decisions and also on a program level have a longer tradition and have been used in one way or another in the past, whereas the use of statistical methods for portfolio decisions is still less familiar in practice (Figure 1).

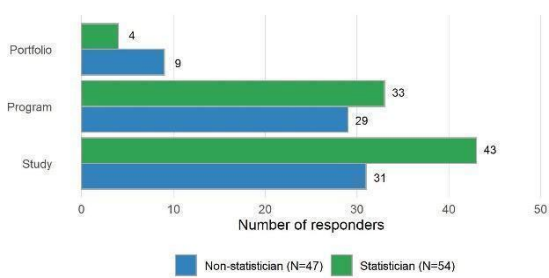

*Figure 1: Overview of the number of survey responses by development level.*

In the following sections, the main results from the three parts of the questionnaire about the use of QDM methods on the trial, program, and portfolio level are summarized. Based on the observed findings, recommendations for improvements are given and actions are suggested to foster the use of these methods in the future.

### 4.1. Trial level

At the trial level, decisions have to be made at the beginning of the trial, e.g., calculation of sample size and power, and at the end of the trial, e.g., success or failure in a confirmatory trial, but also during interim analyses in earlier exploratory development phases.

As there were only few decision makers among the participants, we present the results of the questionnaire grouped by statisticians vs. non-statisticians. The results revealed a high need for decision-making at the trial level; 86% of the statisticians reported that they need to make decisions before the trial, and 95% make decisions during the trial (Figure 2(a)). This is also mirrored by the responses from non-statisticians and decision makers with a higher percentage of decisions before the trial (97%) than during the trial (81%).

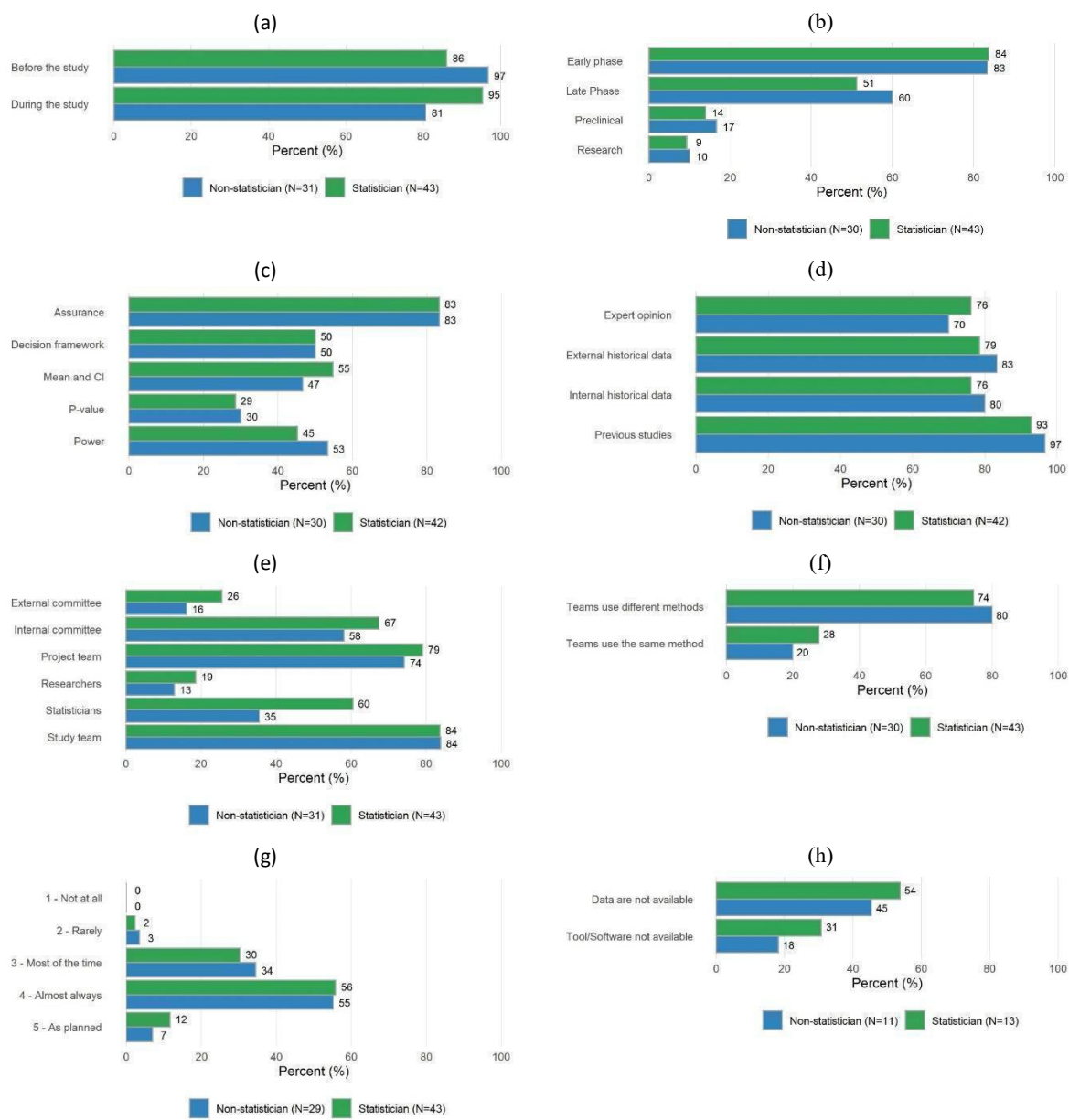

*Figure 2: Decision-making at the trial level. (a) When are study decisions made? (b) At which step of the drug development is quantitative support for decision-making most needed?. (c) Which QDM methods are used at the trial level? CI = confidence interval. Decision framework = e.g. 3-zone approach (Frewer et al. 2016). Assurance = predictive probability of success. Power = conditional probability of success. (d) Which information is used to support decisions on the trial level? (e) Who is involved in trial decisions? (f) Is the same QDM methodology used throughout the company? (g) How strictly are decision rules based on quantitative methods followed? (h) Why are quantitative decision-making methods not used?*

When asked how teams generate the clinical assumptions for their trials, the vast majority of participants reported the use of data from previous studies (93%-97%, Figure 2(d)). In addition, internal and external historical data and publications, and elicitation of expert opinion are very common sources which were selected by approximately 70% to 80% of the responders. Also, combinations of these approaches were mentioned. When more complex information is used, additional comments in the questionnaire revealed that this occurs most often in situations where historical data is only available from a different population or from a surrogate endpoint.

Study decisions are mainly taken directly by the study teams (84%) or project teams (74%-79%) (Figure 2(e)). In addition, internal sponsor committees are involved in trial decisions to a larger extent (58%-67%), whereas external committees play a minor role at this level (16%-26%). In terms of individual roles and functions, statisticians often participate in the decision process, whereas researchers seem to be rarely included. While for the majority of the available answers the results are similar, the two groups have a different perception of the role of the statistician. Only approximately one third (35%) of the non-statisticians answered that statisticians play a role in the decision-making at the trial level. This is in contrast to the view of the statisticians themselves who reported that they participate in trial decisions in the majority of cases (60%).

It is encouraging that quantitative methods are increasingly used for trial planning and trial decisions. The results also revealed that a consistent use of the same methodology across trials is still rare (Figure 2(f)). Only approximately 20% to 30% of the responders confirmed that all trial teams in the organization use the same method. It is more common to use different methods in different projects depending on the situation at hand. In the absence of a common approach defined by the company throughout the organization, the choice of tools and methods depends the trial and program teams.

Another positive finding was that QDM is an important element in the decision-making process if it is used to support trial decisions. In total approximately two thirds of the responders said that prespecified rules are followed (score 5) or almost always followed (score 4) as planned (Figure 2(g)).

The responses provided by the participants show that quantitative decision-making methods are already used in practice. However, there are also obstacles in the way and we tried to elicit the reasons for not using these methods. It became clear that availability of data is the main issue; this was mentioned by approximately half of the responders (45%-54%). But 31% of the statisticians also reported that tools and software are not available (Figure 2(h)).

### 4.2. Program level

The results for the program level were similar to the results at the trial level. However, at the program level, there are less decisions at the beginning of a trial, instead most of the decisions are either at the end of a trial or during trials. In this section, the responses from non-statisticians and decision makers have been combined into one group, labelled as non-statisticians, to allow comparison with the responses from the statisticians. The questionnaire results showed that for program level decision making, similarly to trial level decisions, QDM is mainly used during phase I to phase III studies, and the success criteria are usually developed before a study has begun. Although assurance is the main method used for program decisions (89-94%), as was seen for the trial level decisions, at this level, power is also used in 50-62% of decisions (Figure 3(a)).  The same barriers to using QDM methods as the study level, availability of data and software, were reported at the program level.

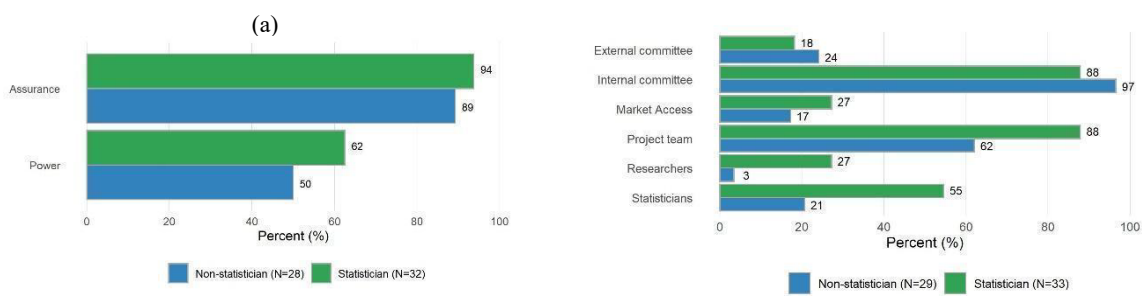

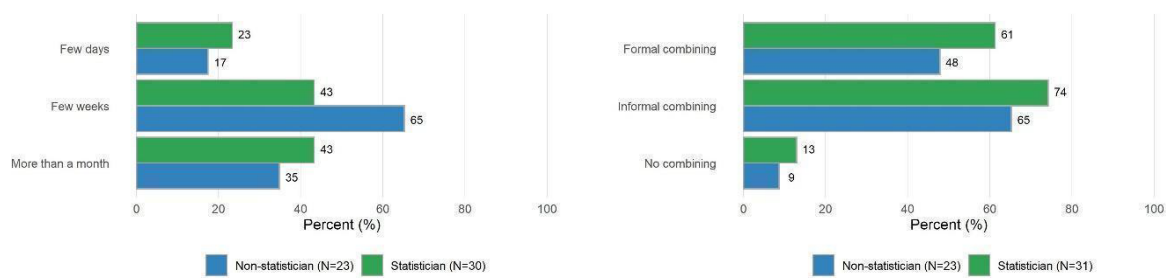

*Figure 3: Decision-making at the program level. (a) Which quantitative decision-making methods are used at the program level? Assurance = predictive probability of success. Power = conditional probability of success. (b) Who is involved in project decisions? (c) How fast are program level decisions made? (d) How are multiple sources of information incorporated to support decisions?*

Project decisions are usually taken by either the project teams and/or internal committees (Figure 3(b)). The market access department is sometimes involved in the decision making at the program level, as are external committees. A key difference between statisticians and non-statisticians is that, statisticians report being involved with project decisions 55% of the time and believe researchers to be involved 27% of the time, whereas the non-statisticians report the statisticians as being involved with project decisions only 21% of the time and researchers only 3% of the time.

Similarly to the trial level results, approximately two thirds (62%) of responders said that prespecified rules are followed or almost followed as planned. Moreover, once the results are available, often it takes a few weeks or more than a month for a decision to be made (Figure 3(c)). This suggests that pre-specified decision criteria support the decision-making process, but program results usually require discussion or confirmation from an internal committee before the decision is regarded as final. Part of this discussion may be around incorporating multiple sources of information (e.g. manufacturing, competitive landscape) as there is no common method for how this is performed. The majority of statisticians apply informal (74%) or formal methods for combining multiple sources of information (Figure 3(d)). The variety of formal and informal approaches utilized add to the variety of methodology used for decision making at the program level.

### 4.3. Portfolio level

Only 13 responses were received for the portfolio level, 4 statisticians and 9 decision makers/non-statisticians, therefore the responses presented in this section combine the responses from the statisticians with the non-statisticians. Although we would expect less people involved at this level, the very low response rate suggests that statisticians are rarely involved with portfolio level analysis or predictions.

Decisions at this level are typically made annually (27%) or on an as needed basis (73%). The decision makers are always internal committees, with some input from project teams (38%) and the market access department (38%) (Figure 4(a)). The increase in the market access department involvement may be expected given the greater financial implication of these decisions.

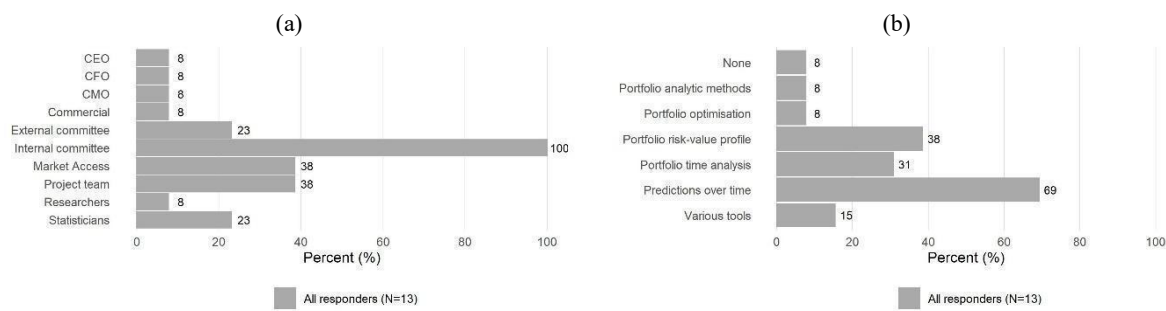

*Figure 4: Decision-making at the portfolio level. Overall summary based on all responders, i.e., based on combined responses from statisticians (4), and non-statisticians and decision-makers (9). (a) Who is involved in portfolio decisions? (b) . Quantitative decision-making methods used at the portfolio level.*

At the portfolio level, success criteria are usually developed (64%). When success criteria are developed, they are often based on financial objectives (100%) or duration/time objectives (86%). Although these criteria are different to the criteria used at trial or program level, quantitative methods are still typically used to support decision making and increase the financial value of the portfolio.

Given the different success criteria at portfolio level, the quantitative methods used are also different to what is used at the trial and program level. The most common quantitative method used is predictions over time (69%), e.g., the chance of obtaining different numbers of marketing authorizations over a pre-specified time horizon. Other methodology includes portfolio risk value profile (38%), e.g. what is the chance of obtaining a variety of net present values, and portfolio time analysis (31%) e.g. investigating the average time for each development phase (Figure 4(b)).

When using the analysis methods, uncertainty about assumptions is usually taken into account (83% of participants), and usually multiple sources of information are used and informally combined.

There is support for increased use of QDM at this level of decision making. Suggestions for further use of QDM by participants included resource allocation and portfolio prioritization, particularly for pre-clinical development. However, the same concerns about availability of data and quality of data leading to unreliable results, as highlighted for the trial level and program level, was also highlighted by these participants.

## 5. Discussion

This paper does not present an exhaustive literature review of existing QDM methods, nor does it claim to be a critical appraisal of QDM methods, presenting their advantages and pitfalls. Instead, the ESIG QDM present the motivation for the use of decision support methods and we focus on examples of good practices which are infusing the statistical and non-statistical community and should be in our view promoted.

The questionnaire created by the ESIG QDM aimed at an informal elicitation of information about the awareness and use of QDM methods among statisticians, non-statisticians and decision makers in the pharmaceutical industry. The results indicate in which development phases quantitative decision-making is used, which methods are applied, which roles or groups are involved in the decision processes, and whether a consistent approach is used within organizations.

In general, there is a high demand for decision making at the trial level, both before and during/after a trial is conducted. Non-statisticians focused more on decisions before the trial, whereas statisticians reported more frequent decisions during/after the trial. We may hypothesize that this reflects the fact that statisticians are more concerned with trial results of interim and final analyses, whereas non-statisticians and decision makers are more involved in Go/NoGo decisions before the start of a trial. It may also indicate that a higher proportion of non-statisticians believe that quantitative work is mostly done by statisticians before the trial starts, e.g., trial design, simulation of design operating characteristics, and sample size calculation, whereas post-trial work for decision support is not visible to the same extent to them.

An interesting finding is the difference in the perception of the role of statisticians in the decision-making process. Non-statisticians and decision makers do not seem to view the statistician as an important decision-maker in the process, as can be seen in Figures 2(e) and 3(b). This outside view contrasts with the self-perception of the statisticians who consider themselves as a major role. A potential explanation could be that statisticians consider themselves as being more involved in decisions because they provide the analysis results which form the basis of the decision making, whereas non-statisticians may focus more on the subsequent decision-making process once the results are delivered and do not feel that statisticians have a major role in this. It may also point to a lack of visibility of statisticians in general and lack of appreciation of the added value decision-making based on statistical principles can bring to pharmaceutical development.

This also ties in with the fact that a consistent approach is still rare in the pharmaceutical industry. A company-wide unified approach should be promoted to foster consistency in the decision-making process across trial teams and to bring the full advantages of quantitative decision-making to bear.

It is also recognized that the availability of data is currently still a limiting factor, and more reliable data is needed to support data-driven decision-making. Another factor hindering the widespread use of quantitative decision-making methods is the availability of tools and software making the application of these approaches easy and efficient. When these obstacles are addressed, trial-level QDM methods can become standard tools in the armory of the working statistician in the future.

At the program level, decisions seem to be predominantly taken before the trial begins. The calculation of assurance and power are the two most reported tools. Such calculations are often done during the planning of phase III trials, and the results show that assurance has gained popularity in recent years but classical power is also still widely used.

The results on the program level shows much consistency with the trial level findings. A notable difference is that although QDM methods are used throughout a program's duration, the committees making program level decisions are more reliant on traditional methods. This may be due to the variety of QDM methodology used for combining data as well as the more complex decision criteria. Further to this, decisions at the program level commonly take a few weeks or more than a month. This suggests that QDM needs further embedding in program level committees to fully utilize its potential. This is likely to be more efficient if the methodology is consistent throughout a company. Improvements to methodology consistency and awareness are likely routes for the decision process to be streamlined resulting in quicker decisions.

The sparseness of responses from statisticians at the portfolio level indicates that statisticians are not commonly involved in strategic decision making. However, given the support for increased use of QDM at this level of decision making, statisticians being more involved would likely facilitate more research into the use of QDM methods for the performance metrics, along with the analysis methods at this level, and this is likely to be encouraged by the decision makers to support decision-making in addition to their judgement and experience .

It should be recognized that the present questionnaire has some limitations. The questionnaire has been proposed to get a picture of the current status of the practical use of QDM methods and does not pretend to be representative of all statisticians and decision-makers population. Also, even if some clear benefits are expected from QDM methods enabling data-driven and evidence-based decisions, a direct comparison between such methods and a more traditional approach based more on intuition/human judgment is hard to make; some retrospective analyses could be envisaged but this may be limited to the trial and potentially the program level decision.

## 6. Conclusions

In recent years the pharmaceutical industry has faced the challenge of increasing costs and constantly high attrition rates which threaten to make its business model unsustainable in the future. To address this challenge and to contribute a possible mitigation, statisticians have proposed the use of QDM methods supporting a data-driven, objective appraisal of the evidence that forms the basis of decisions at different stages of drug development.

Following the need to improve drug development efficiency, QDM has gained traction over the past ten years and is being utilized more widely than before in decision-making processes to enable quicker and more reliable decisions. QDM can add value at all levels of pharmaceutical development, e.g., applying the "quick win or fast fail" approach in early phases. Evidence-based quantitative methods in a decision-making framework provide an integrated assessment of the available evidence and its associated uncertainty thus enabling rational decisions (Ashby 2000). In the last years, traditional frequentist metrics, e.g., type-I error rate and power of hypothesis tests, have been supplemented by the calculation of predictive probability of success (assurance, O'Hagan et al. 2005), or Go/NoGo probabilities in decision-making frameworks (Lalonde et al. 2007, Fisch et al. 2014, Frewer et al. 2016).

The examples presented show that QDM methodology can be applied to a range of decisions within the drug development pathway, from study level decisions to portfolio decisions. Go/NoGo decision making criteria for a phase II oncology trial were used to optimize the phase II trial design in light of the associated risks and thereby make the best use of available resources to show preliminary evidence of efficacy (Frewer et al. 2016). Assurance for a phase III cardiovascular trial was calculated to give the team a more realistic view of the chance of reaching a successful marketing authorization avoiding, e.g., excessive over-confidence after a positive phase II (Crisp et al. 2018). On the portfolio level, a model for dynamically allocating budget to different phase III programs was applied to maximize the expected net present value of a portfolio under the constraint of a given total budget (Patel and Ankolekar 2015).

Growing awareness among statistical leaders in the industry led to the creation of the European EFSPI/PSI special interest group (ESIG) on quantitative decision making in 2017 to share experiences, collect best practices, and promote the use of QDM. The ESIG created a questionnaire in 2019 to learn how and to which extent QDM methods are currently used in the different development phases. The main questionnaire findings were presented in this paper and showed where QDM is already used now, but also some areas for future improvement.

QDM methods are already quite routinely used at study and program levels, but further development is needed at the portfolio level. Despite the increased use of QDM methodology, there are barriers such as the absence of a systematic approach, and the lack of awareness and understanding by decision-makers within an organization. These could be reduced by companies unifying their approach and training their teams in the methodology. Other barriers to utilization of QDM methods include access to data and availability of tools and software to implement these methods. If these practical obstacles could be overcome, QDM methods may become standard tools for the working statistician in the future. Another key aspect is the influence and visibility of statisticians within decision-making teams. Given the expertise statisticians have in quantifying uncertainty, statisticians should increase their visibility, involvement and leadership in cross-functional development training, planning, and decision-making to be seen as leaders in the decision-making process.

The questionnaire revealed, among other things, the different views of the role of the statistician in the decision-making process by statisticians and non-statisticians. The latter group does not seem to perceive statisticians as leaders when decisions are concerned. This indicates that statisticians should increase their visibility and involvement in cross-functional decision-making and is in line with the need to build statistical leadership as recently mentioned by Gibson (2019). Statistical thinking helps to make better decisions by increasing awareness of the risks in the presence of uncertainty and by quantifying the risks involved in the decisions. In this sense, statisticians should embrace QDM methods to better support decisions and eventually to increase the chances of successful drug development. Applying QDM methods in a systematic way can lead to higher success rates, shorter development times, and lower costs (Morgan et al. 2018).

Quantitative decision-making has clear advantages and a number of statistical methods are now available to implement these approaches. Therefore the expert panel of the ESIG on quantitative decision-making encourages their wider use in practice Statisticians should take this opportunity to take leadership and use their skills in quantifying uncertainty and foster data-driven, objective decisions at the different levels of pharmaceutical development.

## 7. Acknowledgements

We thank Pierre Colin for conducting the survey on behalf of the ESIG QDM and for his support in the analysis of the data.